\documentclass[useAMS,usenatbib]{mn2e}
\usepackage{graphicx}
\title[New Wolf-Rayet stars in Circinus]{Discovery of two Galactic Wolf-Rayet stars in Circinus}
\author[A. Roman-Lopes]{A. Roman-Lopes$^{1}$\thanks{roman@dfuls.cl}\\
$^{1}$Physics Department - Universidad de La Serena - Cisternas, 1200 - La Serena - Chile }

\begin{document}

\date{}

\pagerange{\pageref{firstpage}--\pageref{lastpage}} \pubyear{2010}

\maketitle

\label{firstpage}

\begin{abstract}
I report the discovery of two new Galactic Wolf-Rayet stars in Circinus via detection of their C, N and He Near-Infrared emission lines, using
ESO-NTT-SOFI archival data. The H- and K-band spectra of WR67a and WR67b, indicate that they are Wolf-Rayet stars of WN6h and WC8 sub-types, 
respectively. 
WR67a presents a weak-lined spectrum probably reminiscent of young hydrogen rich main-sequence stars such as 
WR25 in Car OB1 and HD97950 in NGC3603.
Indeed, this conclusion is reinforced by the close morphological match of the WR67a H- and K-band spectra with that for WR21a, a known extremely 
massive binary system.
WR67b is probably a non-dusty WC8 Wolf-Rayet star that has a estimated heliocentric distance of 2.7$\pm$0.9 kpc, 
which for its Galactic coordinates, puts the star probably in the near portion of the Scutum-Centaurus arm.

\end{abstract}


\section{ Introduction}

Very massive stars are key actors in the energy balance and chemical evolution of the Galaxy. Their powerful winds, massive 
outflows and expanding H{\sc ii} regions inject large quantity of momentum and energetic ultraviolet (UV) photons into the Galactic interstellar 
medium (ISM). Wolf-Rayet (WR) stars, the chemically evolved descendants of the most massive O type stars, have been suggested as possible
precursors of core-collapse supernovae of types Ib and Ic \citep {b4}. As consequence, they are the main generators of alpha elements 
(such C, N, O, etc) and due to their short life time (considering the age of the Galaxy), they quickly enrich (through supernova explosions) 
the ISM with large quantity of heavy elements \citep{b5}.

%Table 1

\begin{table}
%\begin{minipage}[t]{\columnwidth}
\caption{Small sensus of the known Galactic Wolf-Rayet stars.}
\label{catalog}
%\centering
\renewcommand{\footnoterule}{}  % to avoid a line before footnotes
\begin{tabular}{cccc}
\hline \hline
Reference & Year & Number & Note \\
\hline
    van der Hucht    & 2006  & 298  & 113 WC, 171 WN, \\
                     &       &      &   10 WN/WC, 4 WO \\
    Hadfield et al.  & 2007  & 15   & 4 WC, 11 WN \\
    Kurtev et al.    & 2007  & 4    & WN \\
    Barniske et al.  & 2008  & 2    & WN \\
    Mauerhan et al.  & 2009  & 12   & 3 WC, 9 WN \\
    Shara et al.     & 2009  & 41   & 26 WC, 15 WN \\
    Gvaramadze et al.& 2009  & 1    & WN \\
    Mauerhan et al.  & 2010  & 15   & WN \\
    Gvaramadze et al.& 2010  & 1    & WN \\
    Wachter et al.   & 2010  & 6    & 1 WC, 5 WN \\
    %Total            &       &      &  \\
\hline
\end{tabular}
%\end{minipage}
\end{table} 

%Table 2

\begin{table}
%\begin{minipage}[t]{\columnwidth}
\caption{Summary of the NTT/SofI dataset used in this work.}
\label{catalog}
\centering
\renewcommand{\footnoterule}{}  % to avoid a line before footnotes
\begin{tabular}{cc}
%\hline \hline
%ID & SpType \\
\hline
   Date  & 2005-06-23\\
   Telescope  & NTT\\
   Instrument & SofI\\
   Grism  & GR\\
   Slit  & 0.6$\arcsec$ x 290$\arcsec$\\
   Resolution  & 1000\\
   Coverage ($\mu$m) & 1.53-2.52\\
   Seeing (\arcsec)  & 1-2\\
\hline
\end{tabular}
%\end{minipage}
\end{table} 

%Table 3

\begin{table*}
%\begin{minipage}{226}
\caption{Coordinates and photometric parameters of the newly-discovered WR stars.}
\label{catalog}
\centering
\renewcommand{\footnoterule}{}  % to avoid a line before footnotes
\begin{tabular}{cccccccccccc}
\hline \hline
Object & RA     &  Dec  & $B$ & $V$  & $J$ & $H$ & $K_S$ & $[3.6]$ & $[4.5]$ & $[5.8]$ & $[8.0]$ \\
 &(J2000) & (J2000) &~    &    \\
\hline
   WR67a   &15:16:36.96  & -58:09:58.7  &   16.25  &   14.75  &   9.86  &   9.18 & 8.82 &  8.48  &  8.24  &  8.12  &  7.67 \\
   WR67b   &15:17:46.30  & -57:56:59.2  &   19.35  &   17.25  &   10.34  &   9.26 & 8.46 &  7.90  &  7.56  &  7.37  &  7.09 \\
\hline
\end{tabular}
%\end{minipage}
\end{table*}

%Figure 1

   \begin{figure*}
    %\vspace{302pt}
   \centering
   \includegraphics[bb=14 14 578 173,width=18cm,clip]{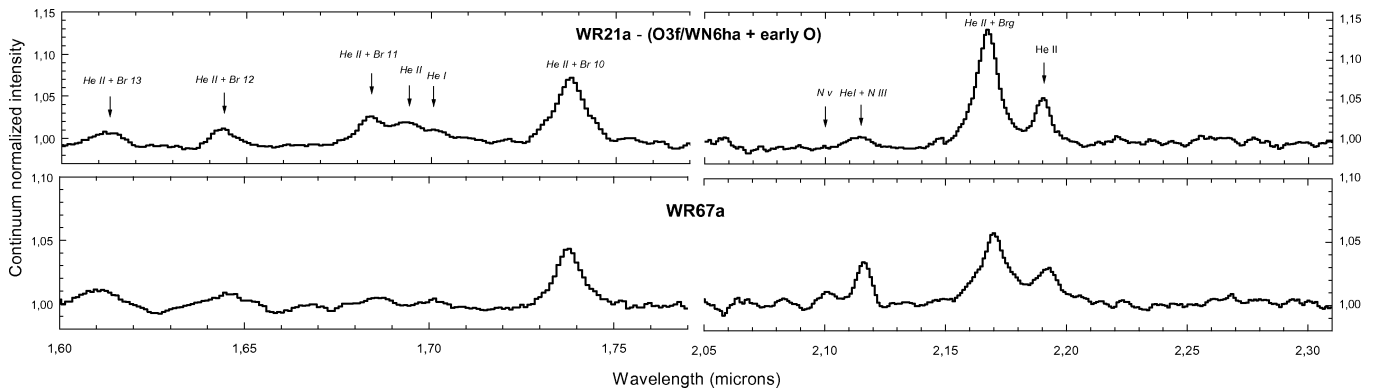}
      \caption{The H- and K-band continuum normalized spectra (bottom) of the WN6h star (WR67a), 
and the ESO-NTT archival spectra (same instrumental setup as for WR67a) for the known O3f/WN6ha + early O very massive (with minimum masses of 
87M$_\odot$ and 53M$_\odot$, respectively) binary system WR21a (Niemela et al. 2008). 
The main H, He and N emission lines are identified by labels. The similarity between the spectra from both sources is remarkable, 
suggesting that WR67a is also a very massive star.}
         \label{Fightback}
   \end{figure*}

Despite the dominant role that WR stars probably play in shaping both, the Galactic structure and its chemical evolution, the total number of 
WR stars and their spatial distribution in the Milk Way is still an open issue. \citet{b6,b7} estimate a number between 1000-2500 WR stars
to be located within the Galaxy. The last census on the Wolf-Rayet Galactic population was made by \citet{b15}, which computed 298
as the known WR stars in the Galaxy. I searched in the literature for new discoveries (see Table 1), and found that up to date, approximately 
100 new Galactic members were discovered. The total number of $\sim$ 395 known Galactic WR stars 
suggest that we still have a large number of such stars to be discovered in the Galaxy.  
One reason for this issue, comes from the fact that dust obscuration makes impossible to observe such stars in the optical window
through the entire Galactic plane. 
In this sense, infrared observations can provide the means for finding a significant fraction of the remaining population of obscured Galactic WR stars. 
Indeed, in the Near-Infrared (NIR) window, H- and K-band spectroscopy may be used to properly identify and classify the new Galactic WR stars, 
accordingly to the known WC, WN and WO subtypes \citep{b1,b2,b3}.

In this paper I report two newly-discovered Wolf-Rayet stars in Circinus via detection of their NIR C, N and He emission lines, using 
ESO-NTT-SOFI archival data.
In section 2 I describe the observational data and the data reduction procedures, in section
3 I present the results, and in section 4 it is presented a summary of the work.

\section{ESO archival data and reduction procedure}

%Figure 2

  \begin{figure*}
    \vspace{10pt}
   \centering
   \includegraphics[bb=14 14 422 124,width=15 cm,clip]{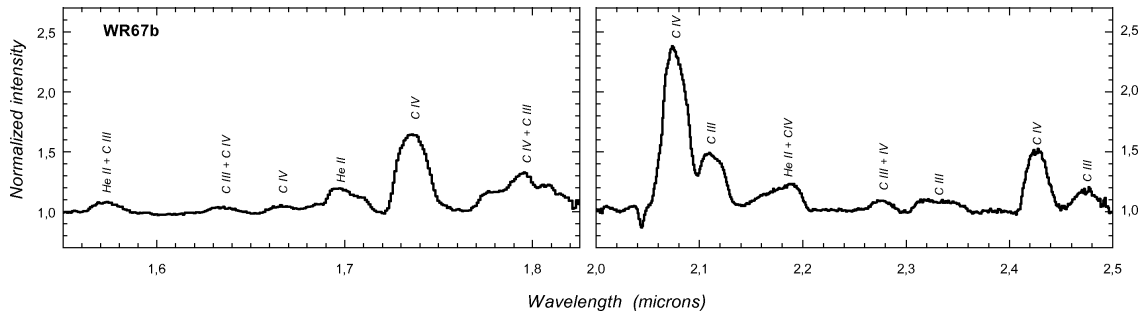}
      \caption{The continuum normalized H- and K-band spectra of the WC8 Wolf-Rayet star WR67b.}
         \label{Fightback}
   \end{figure*} 

In this work I used ESO\footnote{http://archive.eso.org/eso/eso\_archive\_main.html} archival spectroscopic data obtained 
with the SofI instrument \citep{b8}, coupled to the 3.6m NTT telescope.
The spectra were taken as part of the ESO program 075.D-0210(A) (PI A.P. Marston and collaborators), with the targets being selected
accordingly the technique presented by \citet{b4}. A summary of the spectroscopic dataset is presented in Table 2.

The raw frames were reduced following the NIR reduction procedures presented by \citet{b9}, and \textbf{briefly} describe here.
The two-dimensional frames were sky-subtracted for each pair of
images taken at the two nod positions A and B, followed by division
of the resultant image by a master flat. 
The multiple exposures were combined, followed by one-dimensional extraction of the spectra.
Thereafter, wavelength calibration was obtained using the IDENTIFY/DISPCOR IRAF tasks applied to a
set of OH sky line spectra (each with about 35 sky lines in the range 15500A-23000A). The typical error (1-$\sigma$) for this 
calibration process is estimated as $\sim$20~\AA, which corresponds to half of the mean FWHM of the OH lines in the mentioned spectral range.
Telluric atmospheric corrections were done using H- and K-band spectra of B type stars.
The photospheric absorption lines present in the high signal-to-noise telluric spectra, were subtracted from a careful fitting 
(from the use of Voigt and Lorentz profiles) to the hydrogen and helium absorption lines (the He absorption lines are sometimes seem 
at 1.70$\mu$m and 2.11$\mu$m in case of earliest B-type stars), and respective adjacent continuum. This method works fine in case of moderate 
to early B-type stars because the hydrogen lines are not too strong, thus enabling a good determination of the adjacent continuum. 
On the other hand, the strong telluric absorption bands below about 1.55$\mu$m make difficult the use of this method for lines below this wavelength.

%Table 4

\begin{table*}
%\begin{minipage}[t]{\columnwidth}
\caption{List of lines detected in the H- and K-band spectra of WR67a, and the corresponding equivalent  line 
widths.}
\label{catalog}
\centering
\renewcommand{\footnoterule}{}  % to avoid a line before footnotes
\begin{tabular}{cccccccccccc}
\hline \hline
& \textbf{ID} & He{\sc ii}+Br13  & He{\sc ii}+Br12 &He{\sc ii}+Br11 &He{\sc i}  & He{\sc ii}+Br10 & He{\sc i} & N{\sc v} & He{\sc i}+N{\sc iii} & He{\sc ii}+Br$\gamma$ &He{\sc ii} \\
\hline
& \textbf{$\lambda$ ($\micron$)} &1.610 & 1.645  & 1.685 & 1.702 & 1.738 & 2.058 & 2.100 & 2.115 & 2.167 & 2.189 \\
\hline
& EW (\AA{}) & 2.9 $\pm$0.5  & 2.9 $\pm$0.5 & 2.2 $\pm$0.5 & 0.6 $\pm$0.2 & 4.6 $\pm$0.7 & 0.5 $\pm$0.1 & 0.7 $\pm$0.2 & 2.7 $\pm$0.5  & 10.7 $\pm$1.3  & 3.9 $\pm$0.7 \\
\hline
\end{tabular}
%\end{minipage}
\end{table*} 

\section{Results and Discussion}

%Table 5

\begin{table*}
%\begin{minipage}[t]{\columnwidth}
\caption{List of the main lines detected in the H- and K-band spectra of WR67b, and the corresponding equivalent emission line 
widths.}
\label{catalog}
\centering
\renewcommand{\footnoterule}{}  % to avoid a line before footnotes
\begin{tabular}{cccccccccccc}
\hline \hline
& ID & He{\sc i}+He{\sc ii} & C{\sc iv} & C{\sc iii}+C{\sc iv}  & C{\sc iv} & C{\sc iii} & He{\sc ii}+C{\sc iv}  & C{\sc iii}+C{\sc iv} & C{\sc iii} & C{\sc iv} & C{\sc iii} \\
\hline
& $\lambda$ ($\micron$) & 1.698                &     1.736 & 1.801                 & 2.076     & 2.113      & 2.187                 & 2.278                & 2.323      & 2.428     & 2.476 \\
\hline
& EW (\AA{}) & 31$\pm$5              & 214$\pm$25 & 117$\pm$15            & 783$\pm$80 & 234$\pm$25  & 67$\pm$7                & 36$\pm$4              & 35 $\pm$4    & 223 $\pm$25 & 87 $\pm$10 \\
\hline
\end{tabular}
%\end{minipage}
\end{table*}

\subsection{Photometric parameters of the newly-identified Wolf-Rayet stars}

Coordinates and photometric parameters for the newly discovered Wolf-Rayet stars are shown in Table 3. The near- to mid-IR photometric data were 
taken from the NASA/IPAC\footnote{http://irsa.ipac.caltech.edu/applications/BabyGator/} infrared science archive, while the B- and V-band photometry 
were obtained from the Naval Observatory 
Merged Astronomical Dataset (NOMAD\footnote{http://www.nofs.navy.mil/data/fchpix/vo\_nofs.html}).
The stars were named WR67a and WR67b,
following the common practice of to give numbers to Galactic WR stars in RA order, with further additions 
between integers from van der Hucht (2001), being done by adding a, b, c, etc.

\subsection{NIR Spectral Classification}

From the NTT-SofI H- and K-band spectra, I derived spectral types for the two new WR stars,
using the NIR classification scheme proposed by \citet{b12}.

\subsubsection{The WN Wolf-Rayet star}

In Figure 1 I present the telluric corrected (continuum normalized) H- and K-band spectra of the WR67a star, 
together with the H- and K-band ESO archival spectra for WR21a, a well known massive Galactic binary star \citep{b24}.
The strongest features are those corresponding to the blend of H and He emission lines at 1.736$\mu$m and 2.167$\mu$m,
as well as, the He{\sc i}+N{\sc iii} and He{\sc ii} emission lines at 2.115$\mu$m and 2.189$\mu$m, respectively. I can also 
point out the presence of a less intense emission line due to N{\sc v} at 2.100$\mu$m.
From a quickly inspection of the WR67a NIR spectra, one can infer that it is probably a WR of the nitrogen sub-type.
The list with the main detected lines and corresponding equivalent line widths is presented in Table 4.

We can see that WR67a has the N{\sc v} 2.100 $\mu$m emission line in its K-band spectrum, which suggest a subtype earlier 
than WN7 \citep{b12}. 
Also, from the observed ratios of the N{\sc v} (2.100$\mu$m) to He{\sc i}+N{\sc iii} (2.115$\mu$m), and He{\sc ii} (2.189$\mu$m) to 
He{\sc ii}+Br$\gamma$ (2.167$\mu$m) equivalent widths (0.25 and 0.46, respectively), a WN6 subtype could be derived \citep{b12}.
However, strong lined WN6 stars typically possess EW(HeII 2.189) of about 100\AA, while weak lined WN6 stars possess EW(HeII 2.189) 
of about 50\AA (e.g. Crowther \& Smith 1996), versus 3.9\AA, a value an order of magnitude lower than that expected for 
the weak-lined case.

Such kind of weak-lined spectrum as that for WR67a, is reminiscent of young hydrogen rich \textit{main-sequence} WN6h stars such as 
HD93162 (WR25) in Car OB1 and HD97950 in NGC3603, that appear mimicking the spectral appearance
of WR stars, due to their high-luminosities drive dense and fast winds \citep{b26}.
Indeed, this conclusion is reinforced by the close morphological match of the WR67a
H- and K-band spectra with that for WR21a (see Figure 1), an extremely massive binary system (O3f/WN6ha + early O) for which \citet {b24} 
estimated M(WN6)\textgreater87M$_\odot$ and M(O)\textgreater53M$_\odot$.

On the other hand, it is also possible to propose an alternative scenario in which WR67a is in reality a much lower mass weak-lined WN star, 
which has an emission line spectra diluted by the 
presence of a binary/line-of-sight companion. In this sense, further UV and optical spectroscopic survey are necessary to properly address this question.

%Figure3
   \begin{figure}
    %\vspace{302pt}
   \centering
   \includegraphics[bb=14 14 506 780,width=9 cm,clip]{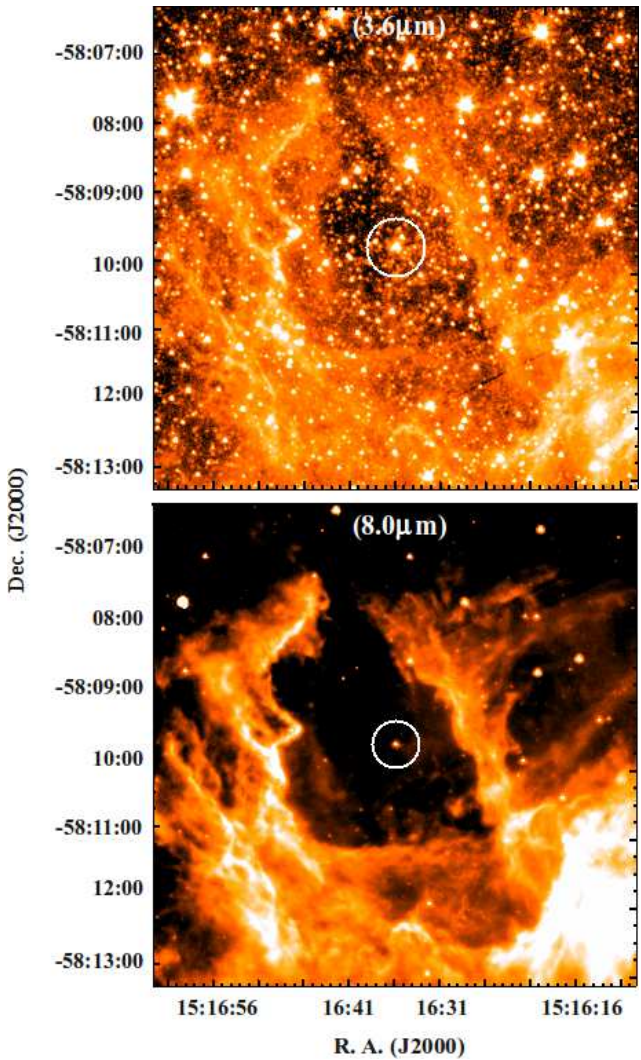}
      \caption{The 7 arcmin $\times$ 7 arcmin GLIMPSE $3.6\mu$m and $8.0\mu$m images of the region around the newly-discovered Wolf-Rayet 
               star WR67a. The star is indicated by \textbf{the white circles}; North is to the top, East to the left. WR67a is the unique point source
near \textbf{the centre of the cavity nicely seem in the} $8.0\mu$m image. \textbf{Indeed}, the hot dust extended emission traces very well the effects of its 
powerful stellar wind into the surrounding nebula. \textbf{A colorized version of this figure is available in the electronic version of the article.}}
         \label{Fightback}
   \end{figure}

\subsubsection{The WC Wolf-Rayet star}

In Figure 2 I present the H- and K-band spectra of the WR67b star, with the
main emission lines indicated by labels. From a qualitative inspection of the spectra, one can immediately conclude that WR67b is a carbon 
type WR star, as  evidenced by the broad and strong Carbon emission lines at 2.076$\mu$m (C{\sc iv}) and 2.113$\mu$m (C{\sc iii}). 
In Table 5 I present a summary of the main detected emission lines together with the corresponding equivalent widths.

Using the observed ratio of the C{\sc iv} (2.076$\mu$m)
and C{\sc iii} (2.113$\mu$m) equivalent widths (Table 5), and the NIR classification scheme for carbon WR stars summarized in Table 6 of \citet{b12}, 
one can conclude that WR67b is a WR star of WC8 subtype.
Late-type WC stars sometimes are surrounded by circumstellar hot dust whose thermal emission may contribute to the underlying NIR continuum.
The observed NIR colours of WR67b are compatible with those for a non-dusty WC8 
type WR star (J-H=0.05, H-K=0.38 - \citet{b12}), reddened by about 7-10 magnitudes in the V band.
Also, the carbon lines detected in the K-band spectrum of WR67b (Figure 2), do
not seem to be "diluted" by continuum emission from circumstellar hot dust or by the presence of a line-of-sight companion. 
However, the absence of reduced line strengths, in principle 
can not be used to discard the presence of a line-of-sight/binary early-type companion since relatively short period late-type WC binaries
like $\gamma$ Vel (WC8+O), are known not to form dust.

\subsubsection{Reddening and distance}

Despite quantitative analysis of the two newly-discovered WR stars is beyond the scope of this work,
from the derived spectral types and related NIR photometry, I computed mean values for reddening and heliocentric distances to 
each new WR star, using the intrinsic colours and absolute magnitudes given by \citet{b12, b30}, considering the extinction law for NIR bands 
derived by \citet{b3}. For WR67a I used the average of the (J-K) intrinsic colours for WR10 (WN5h) and WR78 (WN7h), while the absolute M$_K$ magnitude 
was computed from the average of the values from Table 4 presented by \citet{b30}, considering an associated uncertainty of $\pm$0.5 magnitudes.
The results are shown in Table 6.

For WR67a I computed distances and reddening taking into account the two scenarios mentioned in Section 3.2.1. In 
the first, the WN star is assumed to belongs
to the WN6h sub-type for which M$_K$=-6.95 and (J-K)$_0$=0.06, while in the the second situation a WN6(weak) 
star for which M$_K$=-4.41 and (J-K)$_0$=0.18 is considered.
When WR67a is considered a WN6h star, the resulting
distance values is quite different than that for WR67b, as expected considering the remarkable difference between the luminosities of the
two objects. In fact, while the WR67b star has a computed heliocentric distance of 2.7$\pm$0.9 kpc, which for Galactic coordinates 
G321.4 -0.4, puts it in the near side of the Scutum-Centaurus arm, WR67a (Galactic coordinates G321.1 -0.5) 
has a computed heliocentric distance of 10.5$\pm$2.1 kpc, which places it probably in the other side of the Galaxy, just at the 
far portion of the Scutum-Centaurus arm.
On the other hand, in a scenario where one consider WR67a as a weak-lined WN star in which its emission line spectra seems diluted by the 
presence of a binary/line-of-sight companion, a much lower heliocentric distance of 3.4$\pm0.8$ kpc is found.

%Table 6

\begin{table*}
%\begin{minipage}{226}
\caption{Derived extinction and distances together with the $J-K_S$ and $H-K_S$ intrinsic colours and absolute $K$ magnitudes for the 
newly-discovered WR stars(see text for details).}
\label{catalog}
\centering
\renewcommand{\footnoterule}{}  % to avoid a line before footnotes
\begin{tabular}{cccccc}
\hline \hline
Object    & Spec.Type & $M_K$  & $(J-K_S)_0$ & $A_K$ &  d (kpc) \\
\hline
   WR67a  &   WN6h  & -6.95$\pm$0.50  &    0.06 &  0.66$\pm$0.15 &  10.5$\pm$2.1    \\
          &   WN6(weak)  & -4.41$\pm$0.50  &    0.18 &  0.58$\pm$0.14 &  3.4$\pm$0.8 \\
\hline
   WR67b  &   WC8   & -4.65$\pm$0.50          &    0.43         &  1.04$\pm$0.25          &  2.7$\pm$0.9    \\
\hline
\end{tabular}
%\end{minipage}
\end{table*} 

I searched for stellar clusters close to WR67a and WR67b, and found that they are located at about 10 arcsec and 15 arcmin, respectively, from the 
centre of the cluster candidate MCM2005b \#60 \citep{b11}, 
that has central coordinates $\alpha$=15:16:36 and $\delta$=-58:10:07 (J2000), which however is not obviously seem in the 2MASS K$_S$-band image.
However, from the GLIMPSE 3.6$\mu$m and 8.0$\mu$m images of the
field around WR67a (Figure 3), one can notice the presence of several point like sources (mainly in the 3.6$\mu$m image) inside
a cavity delineated by the hot dust extended emission, with WR67a being one of the most prominent. Also from the 
GLIMPSE $8.0\mu$m image of the region around the newly-discovered WN star, it can be seem the hot dust extended emission nicely tracing 
what seems to be the effects of a powerful stellar wind into the surrounding nebula.
Indeed, WR67a appears as the unique point source placed near the centre of the cavity that is seem in the $8.0\mu$m image, which
is probably associated to a Galactic H{\sc ii} region detected by the H 109$\alpha$/H 110$\alpha$ radio recombination line survey 
performed by \citet{b29}.
Their source \#143 (G321.1 -0.55), measuring about 3 arcmin $\times$ 3 arcmin, has coordinates 15:16:36.3 -58:12:10 (J2000). It presents a flux density 
of 8.5 Jy at 5 GHz, and a radial velocity of -56 km/s, measured from the associated hydrogen recombination lines.
From the use of the Galactic rotation curve, they computed far and near heliocentric distances of 11.6 kpc and 3.9 kpc, respectively. Interesting, both 
values are similar to those obtained in this work when WR67a is considered to be a WN6h (10.5$\pm$2.1 kpc) and a WN6(weak) (3.4$\pm0.8$ kpc) type WR star, 
respectively.

\section{Summary}

In this work I report the discovery of two new Galactic Wolf-Rayet stars in the direction of Circinus.
The analysis of their ESO-NTT-SOFI H- and K-band spectra, indicate that WR67a and WR67b are WR stars belonging to the WN and WC types, 
respectively.

WR67a presents a weak-lined spectrum reminiscent of young hydrogen rich \textit{main-sequence} WN6h stars such as 
HD93162 (WR25) in Car OB1 and HD97950 in NGC3603, 
which due to their high-luminosities drive dense and fast winds, appear mimicking the spectral appearance of WR stars.
In one hand, this assumption is reinforced by the close morphological match of the WR67a H- and K-band spectra with that for WR21a, an extremely 
massive binary system (O3f/WN6ha + early O) for which \citet {b24} estimated M(WN6)\textgreater87M$_\odot$ and M(O)\textgreater53M$_\odot$.
In this case, an heliocentric distance of 10.5$\pm$2.1 kpc is computed, placing WR67a in the other side of the Galaxy, 
probably in the far portion of the Scutum-Centaurus arm.
On the other hand, if we consider an alternative scenario where WR67a is supposed to be a weak-lined WN star showing an emission line spectra 
diluted by the presence of a binary/line-of-sight companion, a much lower heliocentric distance of 3.4$\pm0.8$ kpc is found.

WR67b is a WC8 Wolf-Rayet star probably placed at an heliocentric distance of 2.7$\pm$0.9 kpc, 
which for Galactic coordinates G321.4 -0.4, puts it probably at the near portion of the Scutum-Centaurus arm.
Late-type WC stars sometimes are surrounded by circumstellar hot dust, but
the observed NIR colours of WR67b seems to be compatible with those for a non-dusty WC8 type WR star reddened by about 7-10 magnitudes in the V band.
Indeed, its K-band carbon lines do not seem to be "diluted" by the presence of a line-of-sight companion. 
However, the absence of reduced line strengths, in principle can not be used to discard the presence of a line-of-sight/binary early-type 
companion since relatively short period late-type WC binaries (e.g. like $\gamma$ Vel (WC8+O)), are known not to form dust.

\section*{Acknowledgments}

      I would like to thanks the comments and suggestions made by the referee, Dr. Paul Crowther. His comments were 
      very useful to improve the quality and presentation of the final manuscript. 
      This work was partially supported by the ALMA-CONICYT Fund, under the project number 31060004,
      "A New Astronomer for the Astrophysics Group, Universidad de La Serena", and by the Physics department of the 
      Universidad de La Serena.
      This research has made use of the NASA/ IPAC Infrared Science Archive, which is operated by the Jet Propulsion Laboratory, 
      California Institute of Technology, under contract with the National Aeronautics and Space Administration.
      This publication makes use of data products from the Two Micron All Sky Survey, which is a joint project of the University of 
      Massachusetts and the Infrared Processing and Analysis Center/California Institute of Technology, funded by the
      National Aeronautics and Space Administration and the National Science Foundation.
      This research has made use of the SIMBAD database, operated at CDS, Strasbourg, France.
      Based on observations made with ESO Telescopes at the La Silla Observatory under program ID 075.D-0210(A).

\end{document}